\documentclass[12pt]{article}
\usepackage{latexsym,amsfonts}
\newcommand{\rr}{{\mathbb{R}}}
\newcommand{\cc}{{\mathbb{C}}}
\newcommand{\sll}{SL(2,$\rr$)}
\newcommand{\slu}{SL(2,$\rr$)/U(1)}
\renewcommand{\d}{{\rm d}}
\newcommand{\e}{{\rm e}}
\renewcommand{\i}{{\rm i}}
\newcommand{\del}{\partial}
\newcommand{\delz}{\partial_z}
\newcommand{\ub}{{\bar{u}}}
\newcommand{\yb}{{\bar{y}}}
\newcommand{\zs}{{\tilde z}}
\newcommand{\zb}{{\bar{z}}}
\newcommand{\delzb}{\partial_\zb}

\newcommand{\delsp}{\partial_{\sigma'}}
\newcommand{\thr}[1]{\tanh^{#1}\!r\;}
\newcommand{\sh}[2]{\sinh^{#1}\!{#2}\;}
\newcommand{\shr}[1]{\sh{#1}{r}}
\newcommand{\ch}[2]{\cosh^{#1}\!{#2}\;}
\newcommand{\chr}[1]{\ch{#1}{r}}
\newcommand{\Ab}{\bar{A}}
\newcommand{\Bb}{\bar{B}}
\newcommand{\psib}{{\bar{\psi}}}
\newcommand{\phib}{{\bar{\phi}}}
\newcommand{\chib}{{\bar{\chi}}}
\newcommand{\halb}{\frac{1}{2}}
\newcommand{\tr}{{\rm tr}}
\newcommand{\intl}{\int\limits}

\newcommand{\vtel}{\frac{1}{4}}
\newcommand{\inv}{^{-1}}
\newcommand{\Lie}{\mbox{Lie}}
\newcommand{\Det}{{\rm Det}}
\newcommand{\GL}{{\rm GL}}
\newcommand{\comm}[2]{\left[#1,#2\right]}
\newcommand{\pk}[2]{\left\{#1,#2\right\}}
\def\matrix#1#2{\left(\begin{array}{#1}#2\end{array}\right)}
\newcommand{\Cb}{{\bar{C}}}
\newcommand{\Db}{{\bar{D}}}

\def\1ad{\mbox{\normalsize $^1$}}
\def\2ad{\mbox{\normalsize $^2$}}
\def\makefront{\vspace*{1cm}\begin{center}
\def\newtitleline{\\ \vskip 5pt}{\Large\bf\titleline}\\
\vskip 1truecm{\large\bf\authors}\\
\vskip 5truemm
\addresses
\end{center}
\vskip 1truecm{\bf Abstract:}
\abstracttext
\vskip 1truecm}
\begin{document}
\def\titleline{The Complete Solution of the Classical \slu{} 
Gauged WZNW Field Theory\makebox[0cm][c]
{\raisebox{3cm}[0cm][0cm]{\parbox{6cm}
{\normalsize DESY 98--062\\MZ--TH/98--21\\
hep-th/9805215}
}}}
\def\authors{Uwe M\"uller \1ad , Gerhard Weigt \2ad}
\def\addresses{
\1ad
Institut f\"ur Physik, Johannes-Gutenberg-Universit\"at Mainz\\
Staudingerweg 7, D-55099 Mainz, Germany\\
\2ad
Deutsches Elektronen-Synchrotron DESY Zeuthen\\ 
Platanenallee 6, D--15738 Zeuthen, Germany\\
E-mail:\ \ umueller@thep.physik.uni-mainz.de,\ \ weigt@ifh.de}

\def\abstracttext{We prove that any gauged WZNW model has a Lax pair
  representation, and give explicitly the general solution of the
  classical equations of motion of the \sll/U(1) theory.  We calculate
  the symplectic structure of this solution by solving a differential
  equation of the Gelfand-Dikii type with initial state conditions at
  infinity, and transform the canonical physical fields non-locally
  onto canonical free fields.  The results will, finally, be collected
  in a local B\"acklund transformation.  These calculations prepare
  the theory for an exact canonical quantization.  }

\makefront

\hfill\parbox{4,5cm}{\footnotesize\sl
  God does not care about\\
  our mathematical difficulties.\\
  He integrates empirically.\\[1ex]
  \hspace*{\fill}Albert Einstein}

\section{Introduction}
To get space-time structure from conformal Wess-Zumino-Novikov-Witten
(WZNW) theories is an especially tempting but mathematically
non-trivial problem.  The standard procedure integrates out
functionally the gauge field of a gauged WZNW model and attempts a
sigma-model interpretation of the arising effective action.  An
outstanding example is the \slu{} coset theory which gives a
space-time black hole metric \cite{Witten}.  However, that procedure
neither preserves conformal invariance in a convincing manner
\cite{Witten,DVV,Tseytlin} nor delivers an exact effective action
\cite{Buscher} for dynamical calculations. The path-integration left
uncalculated a functional determinant of a differential operator
\cite{Mu} which might influence previous results.

Therefore, it would be desirable to have a mathematically well-defined
method. In this paper we take up again the discussion of the \slu{}
gauged WZNW theory and follow an entirely classical approach based on
an exact effective action \cite{BA} obtained by eliminating the
non-dynamical U(1) gauge field algebraically instead of
path-integrating it out. The arising coset theory is conformal and
integrable \cite{MuWe}. More generally, we prove that integrability
holds for any gauged WZNW theory by deriving a Lax pair for its
equation of motion. So far this was known only for nilpotent gauging
which yields Toda theories \cite{Balog, GS}. We do not look here for
algebraic coset current constructions \cite{GKO} but present the
general analytic solution of the classical equations of motion of the
\slu{} theory, and we derive the symplectic structure for a
field-theoretic situation by solving a Gelfand-Dikii type second order
differential equation with boundary conditions at infinity. The
likewise solved problem with periodic boundary conditions is more
involved due to zero modes. This solution which describes a closed
string moving in the background of a black hole target-space metric
will be prepared for canonical quantization in a separate paper.
Canonical quantization of the \slu{} model should be possible once we
have derived from the symplectic structure canonical transformations
of the physical fields onto free fields.  These transformations are
non-local, but this non-locality can be hidden superficially in a
local B\"acklund transformation \cite{MuWe}.

Since all considerations are classical there will be no dilaton here
which usually arises in pertubative $\beta$-function calculations due
to the curved target-space metric of the model \cite{Callan}.  We hope
that quantization based on the given exact analytical solution will
help to clarify some of the discussions \cite{Witten,DVV,Tseytlin}
related to a possibly non-perturbative dilaton. Black hole radiation
could also appear in a new light once we have quantized the theory.
However, it seems to be reasonable to consider this Lagrangean coset
model in its own right as an integrable conformal field theory with
space-time structure. First results were published in 
ref.\ \cite{MuWe}.

We begin our calculations by first proving that any $G/H$ gauged WZNW
theory is integrable. It will be shown that the non-linear equations
of motion can be expressed by a linear Lax pair representation.

\section{The Integrability of the G/H Gauged\\ 
WZNW Theory}
Wess-Zumino-Novikov-Witten models \cite{WZNW} are $\sigma$-models
\begin{equation}\label{WZNW-Wirkung}
S_{\rm{WZNW}}[g]=\frac{k}{8\pi}\intl_M \gamma^{\mu\nu}
\tr\left(g^{-1}\del_\mu gg^{-1}\del_\nu g\right)
\sqrt{-\gamma}\d^2\xi+kI_{\rm{WZ}}[g]
\end{equation}
which include the additional topological Wess-Zumino term
\begin{equation}\label{WZ-term}
  I_{\rm{WZ}}=\frac{1}{12\pi}\intl_{B,\;\partial B=M}\tr\left( g^{-1}\d g\wedge g^{-1}\d
  g\wedge g^{-1}\d g\right)
\end{equation}
integrated over a volume $B$ with the boundary ${\partial B=M}$. They
describe conformal and integrable theories which are globally
invariant under
left and right multiplications with elements of the symmetry group $G$
\begin{equation}\label{li-re-symm}
S_{\rm{WZNW}}[g]=S_{\rm{WZNW}}[a^{-1}gb],\qquad a,b\in G, \qquad
a,b={\rm const}.
\end{equation}
The field $g(\tau,\sigma)$ takes values in the group $G$,
$\gamma_{\mu\nu}$ is the Minkowskian metric of the world surface $M$
parametrized by $\xi = \{\tau,\sigma\}$, $\gamma$ its determinant, and
$k$ is the coupling parameter of the theory.

In order to get from (\ref{WZNW-Wirkung}) the $G/H$ gauged action, we have
to embed the subgroup $H$ of $G$ into the symmetry group 
$G_{\rm L}\times G_{\rm R}$ of left and right multiplications by
homomorphisms
\begin{eqnarray}\label{LR-homo}
L:H\to G_{\rm L},\quad h\mapsto L(h),\nonumber\\
R:H\to G_{\rm R},\quad h\mapsto R(h).
\end{eqnarray}
Then the gauge transformations corresponding to (\ref{li-re-symm}) become
\begin{equation}\label{eichtrafi}
g\to L(h^{-1})gR(h),\quad h\in H.
\end{equation}
The homomorphic mappings $L$ and $R$ generate linear mappings $L'$ und
$R'$ of the Lie algebra of $H$ into the Lie algebra of $G$ 
\begin{eqnarray}
  \label{LR-Lie-abb}
  &&L':\Lie(H)\to\Lie(G),\quad
  T^{i}\mapsto L'(T^{i})=\left.\frac{\d}{\d t}L
  \left(\exp(T^{i}t)\right)\right|_{t=0},\nonumber\\
  &&R':\Lie(H)\to\Lie(G),\quad
  T^{i}\mapsto R'(T^{i})=\left.\frac{\d}{\d t}R
  \left(\exp(T^{i}t)\right)\right|_{t=0},
\end{eqnarray}
where $T^{i}$ are the elements of the Lie algebra of $H$. Then the
infinitesimal gauge transformations of (\ref{eichtrafi}) become
\begin{equation}\label{eichtrafo-g-inf}
\delta g=gR'(\delta h)-L'(\delta h)g.
\end{equation}
Usually the invariance under local gauge transformations 
$\delta h(\tau, \sigma)$ is found by introducing an $H$ gauge field
$A_\mu$ with
the transformation properties
\begin{equation}\label{eichtrafo-A-endl}
A_\mu\to h^{-1}A_\mu h-h^{-1}\del_\mu h
\end{equation}
or infinitesimally
\begin{equation}
  \delta A_\mu=-\del_\mu\delta h+\comm{A_\mu}{\delta h},
\end{equation}
and replacing partial derivatives by covariant ones
\begin{equation}\label{kov-abl}
\del_\mu g\to D_\mu g\equiv\del_\mu g-L'(A_\mu)g+gR'(A_\mu).
\end{equation}
But this does not guarantee the local gauge invariance of the WZ term
(\ref{WZ-term}) unless the anomaly cancellation condition
\begin{equation}\label{WZ-Eichbed}
\tr\left(L'\otimes L'-R'\otimes R'\right)=0
\end{equation}
is fulfilled. One could specialize this condition to a nilpotent gauge
\begin{equation}
  \label{nilpotenz}
 L'\otimes L'=0,\quad R'\otimes R'=0,
\end{equation}
which reduces WZNW models to Toda theories. One could also choose a
vector gauge
\begin{equation}\label{vektoreichung}
L(h)=R(h)\quad\Rightarrow\quad L'=R',
\end{equation}
or for abelian gauge groups the axial gauge
\begin{equation}\label{axialeichung}
L(h)=R(h^{-1})\quad\Rightarrow\quad L'=-R'.
\end{equation}
Without any specialization, after some manipulations, the $G/H$
gauged\\ WZNW model written in light cone coordinates $z=\tau+\sigma$,
$\zb=\tau-\sigma$ becomes
\begin{eqnarray}\label{WZNW-Wirk-allg-lk}
S_{\rm WZNW, gauged}[g,A]&=&S_{\rm WZNW}[g]+\frac{k}{2\pi}\intl_M
\d z\d\zb\times\nonumber\\
&&\times\Big[
\tr\left(R'(A_z)g\inv\delzb g\right)-
\tr\left(L'(A_\zb)\delz gg\inv\right)-\nonumber\\
&&\phantom{\times\Big[}-
\tr\left(R'(A_z)g\inv L'(A_\zb)g\right)+
\tr\left(R'(A_z)R'(A_\zb)\right)\Big].\nonumber\\
\end{eqnarray}

The local gauge invariance of this action remains valid if the gauge
field $A_\mu$, which behaves non-dynamically here and is, therefore,
purely algebraically determined by its field equation, is eliminated by 
this equation. Then the equation of motion for the field $g(\tau,\sigma)$
follows by varying this action with respect to $g$ only. The variation
of $A_\mu$ as a function of $g$ does not contribute due to the
extremal pinciple.  With the definitions
\begin{equation}\label{B-def}
B\equiv R'(A_z),\quad\Bb\equiv g\inv L'(A_\zb)g
\end{equation}
the equation of motion of $g(\tau,\sigma)$ becomes
\begin{equation}\label{eom-g}
\delz\left(g\inv\delzb g\right)-\comm{B}{g\inv\delzb g}
-\delz\Bb+\delzb B+\comm{B}{\Bb}=0.
\end{equation}
These equations can be rewritten as a flatness condition 
\begin{equation}\label{laxpaar-diffop}
\comm{\delz-B}{g\inv\delzb g}
+\comm{\delz-B}{\delzb-\Bb}=
\comm{\delz-B}{\delzb+g\inv\delzb g-\Bb}=0,
\end{equation}
which is the Lax pair representation we are looking for.  The
commutativity of the linear Lax operators $\left(\delz-B\right)$ and
$\left(\delzb+g\inv\delzb g-\Bb\right)$ is the necessary condition in
order that the system of linear first order differential equations
\begin{equation}\label{laxpaar}
\left(\delz-B\right)\Psi=0,\quad\left(\delzb+g\inv\delzb
g-\Bb\right)\Psi=0
\end{equation}
has a non-trivial solution. This proves the integrability of $G/H$
gauged WZNW models.

We should mention here that the equation of motion (\ref{eom-g}) is an
equation between elements of the Lie algebra $G$. It describes in
components $\dim G$ equations for $\dim G-\dim H$ fields. This becomes
manifest by gauge transforming  the gauge covariant form of (\ref{eom-g})
\begin{equation}\label{eom-g-kov}
D_z\left(g\inv D_\zb g\right)-R'(F_{z\zb})=0
\end{equation}
into
\begin{equation}\label{eichtrafo-eom-g}
R(h\inv)\left(D_z\left(g\inv D_\zb g\right)-R'(F_{z\zb})\right)R(h)=0,
\end{equation}
where
\begin{equation}\label{eichfeldstaerke}
F_{\mu\nu}\equiv\del_\mu A_\nu-\del_\nu A_\mu-\comm{A_\mu}{A_\nu}
\end{equation}
is the field strength with the gauge transformations
\begin{equation}\label{eichtrafo-F}
F_{\mu\nu}\to h\inv F_{\mu\nu}h.
\end{equation}
Parametrizing the group elements $g$ analogously to a Gau{\ss}ian
decomposition
\begin{equation}\label{eich-param}
g(x^\alpha,y^a)=L(h(y^a))\inv g_0(x^\alpha)R(h(y^a)),\quad
\begin{array}[t]{l}
\alpha=1,\ldots,\dim G-\dim H,\\
a=1,\ldots,\dim H,
\end{array}
\end{equation}
the equations of motions for the fields $x^\alpha$, $y^a$ take the form
\begin{equation}\label{eom-xy}
R(h(y^a))\inv
\left(D_z\left(g_0(x^\alpha)\inv D_\zb g_0(x^\alpha)\right)-
R'(F_{z\zb}[g_0(x^\alpha)])\right)R(h(y^a))=0.
\end{equation}
Indeed, due to gauge invariance we have equations of motion for the
fields $x^\alpha$ only.  But we can also derive the $\dim H$
identities among the $\dim G$ components of the equation of motion by
varying the action (\ref{WZNW-Wirk-allg-lk}) with respect to the pure
gauge transformations (\ref{eichtrafo-g-inf}), and using the
decomposition 
\begin{equation}\label{Eichfeld-komp}
A_z=A_z^aH^a,\quad A_\zb=A_\zb^aH^a
\end{equation}
as well as the mapping (\ref{LR-Lie-abb}) in the form
\begin{equation}\label{RL-Abb}
R'(H^a)=R^a,\quad L'(H^a)=L^a,\quad R^a,L^a\in\Lie(G).
\end{equation}
Thus there are $\dim H$ identities
\begin{equation}\label{eom-g-ident}
\tr\left(\left(D_z\left(g\inv D_\zb g\right)-
R'(F_{z\zb})\right)\left(R^a-g\inv L^ag\right)\right)=0.
\end{equation}

The Lax pair representation (\ref{laxpaar-diffop}) is likewise
overdetermined, and it can be reduced as well by equations
(\ref{eom-g-ident}) to an independent set of conditions.

In the following we shall restrict ourselves exclusively to the \sll/U(1)
gauged WZNW theory.

\section{The Classical Effective Action of the\\ 
\slu{} Gauged WZNW Theory}

We parametrize the group SL(2,$\rr$) in accordance with
(\ref{eich-param}) by 
\begin{eqnarray}\label{para}
g&=&g(r,t,\alpha)=\exp\left((2\alpha-t)I/2\right)
\exp(rJ)\exp\left((2\alpha+t)I/2\right)\nonumber\\
&=&
\matrix{cc}{
\cosh r\cos 2\alpha+\sinh r\cos t&\cosh r\sin 2\alpha+\sinh r\sin t\\
-\cosh r\sin 2\alpha+\sinh r\sin t&\cosh r\cos 2\alpha-\sinh r\cos t},
\nonumber\\
&&0\le r<\infty,\quad
0\le t< 2\pi,\quad
0\le \alpha<\pi,
\end{eqnarray}
where the Lie algebra elements $I$ and $J$ are the matrices
\begin{equation}\label{IJ-def}
I=\matrix{rr}{0&1\\-1&0},\quad
J=\matrix{rr}{1&0\\0&-1},
\end{equation}
and gauge the subgroup $U(1)$ with group elements $h$ and generator $\i$
\begin{equation}\label{U(1)}
h(\alpha)=\exp\left(\i\alpha\right),\quad\alpha\in\rr
\end{equation}
according to the recipe (\ref{LR-homo}), (\ref{LR-Lie-abb})
\begin{equation}\label{LR-SLU}
L,R:H\to G,\quad
h=\exp\left(\i\alpha\right)\mapsto L(h)=\exp\left(-\alpha I\right),
R(h)=\exp\left(\alpha I\right)
\end{equation}
in the axial gauge 
\begin{equation}\label{U(1):L'R'}
-L^1=-L'(H^1)=R^1=R'(H^1)=I.
\end{equation}
Then (\ref{WZNW-Wirk-allg-lk}) becomes the \slu{} gauged WZNW action
\begin{eqnarray}\label{W-mit-Eichfeld}
S_{\rm WZNW,gauged}[r,t,\alpha,A]&=&
\frac{k}{2\pi}\intl_M\d z\d\zb\Bigg(\delz r\delzb r+\thr{2}\delz
t\delzb t-
\\
&&\hspace{-4cm}-4\chr{2}\left(A^1_z+\delz\alpha+\halb\thr{2}\delz
t\right)
\left(A^1_\zb+\delzb\alpha-\halb\thr{2}\delzb t\right)\Bigg),\nonumber
\end{eqnarray}
which can be considered as a Lagrangean formulation of the algebraic
coset current construction \cite{GKO}, in case, we make use of the
gauge field $A_\mu$ as a Lagrangean multiplier. This is possible
because, as was mentioned before, the gauge field is non-dynamical.
It is purely algebraically determined by the field equations
\begin{equation}\label{SLU-eom-A}
A^1_z=-\delz\alpha-\halb\thr{2}\delz t\quad\mbox{and}\quad
A^1_\zb=-\delzb\alpha+\halb\thr{2}\delzb t
\end{equation}
which imply in connection with the equations of motions of the fields $r,t$
a vanishing field strength \cite{BA}
\begin{equation}\label{A-flach}
F^1_{z\zb}=\delz A^1_\zb-\delzb A^1_z=0.
\end{equation}
Since the field $\alpha$ of (\ref{para}) behaves in accordance with
(\ref{eom-xy}) as well non-dy\-na\-mi\-cal\-ly, we choose the gauge $\alpha=0$.

These considerations allow us to define, entirely classically, the
effective action of the \slu{} gauged WZNW theory in a gauge invariant
manner by eliminating the gauge field through the algebraic equation
(\ref{SLU-eom-A}) \cite{BA}.  The result
\begin{equation}\label{SLU-Wirk-ohne-A}
S_{\rm WZNW,gauged}[r,t]=
\frac{k}{2\pi}\intl_M\d z\d\zb
\left(\delz r\delzb r+\thr{2}\delz t\delzb t\right)
\end{equation}
is conformal and it describes an integrable theory. It yields, as we
know, the same equations of motion for the fields $r$, $t$ as the
action (\ref{W-mit-Eichfeld}). We could also get this action by
formally Gau{\ss}-integrating over the gauge field and while doing so
neglecting possible anomalies.  The more reasonable path-integration
over the gauge field decomposed in terms of scalar fields $\phi$,
$\chi$
\begin{equation}\label{A-zerl}
  A_\mu=\del_\mu\phi+\epsilon_\mu{}^\nu\del_\nu\chi
\end{equation}
yields quantum contributions to the effective action, but it is
incomplete as well \cite{Buscher} because the functional determinant
$\Det(M)$ of the operator \cite{Mu}
\begin{equation}
  \label{Mdef2a}
  \quad M=-\matrix{cc}{ \del^2+\del^\mu
    g\del_\mu&\epsilon^{\mu\nu}\del_\mu g\del_\nu\\ 
    -\epsilon^{\mu\nu}\del_\mu g\del_\nu&\del^2+\del^\mu g\del_\mu},
  \quad\quad  g=\ln\chr{2}+\ln\frac{k}{\pi}.
\end{equation}
resists exact calculation so far. 

The classical action (\ref{SLU-Wirk-ohne-A}) is, therefore, a reliable
basis for an exact discussion of the \slu{} theory, and we hope that
it is a useful one for quantization. It also has a nice $\sigma$-model
interpretation \cite{Witten}: the target-space metric
\begin{equation}
  \label{line-el-Mink}
  \d s^2=\d r^2+\thr{2}\d t^2
\end{equation}
shows in light cone coordinates 
\begin{equation}
  \label{lichtkegel}
  u=-\shr{}\e^{\i t},\quad
  \bar u=\shr{}\e^{-\i t}
\end{equation}
after Wick rotation $t\to\i t$ a two-dimensional black hole singularity
\begin{equation}
  \label{line-el-Mink3}
  \d s^2=-\frac{\d u\d \bar u}{1-u\bar u}
\end{equation}
with singular curvature tensor. And as a conformal theory the action
describes a string moving in the background of a black hole.

\section{Equations of Motion, Conservation Laws and the 
Gelfand-Dikii Equation}

The equations of motion of the action (\ref{SLU-Wirk-ohne-A}) (or 
(\ref{W-mit-Eichfeld}))
\begin{eqnarray}\label{eom}
  \delz\delzb r&=&\frac{\shr{}}{\chr{3}}\delz t\;\delzb t ,\nonumber\\
  \delz\delzb t&=&-\frac{1}{\shr{}\chr{}} \left(\delz r\;\delzb
  t+\delz t\;\delzb r\right).
\end{eqnarray}
guarantee conservation of the chiral component of the
energy-momentum tensor (we will not indicate, whenever possible,
the similar anti-chiral parts) ($\gamma=\sqrt{2\pi/k}$)
\begin{equation}\label{emt}
  T\equiv T_{zz}=\frac{1}{\gamma^2}\left((\delz r)^2+\thr{2}(\delz
  t)^2\right),
\end{equation}
and vanishing trace
\begin{equation}
T_{z\zb}+T_{\zb z}=0
\end{equation}
shows the conformal invariance of the theory.

Moreover, by multiplying currents of the ungauged \slu{} WZNW model
with Wilson-line factors \cite{BA} there arise two further conserved
chiral quantities on shell
\begin{equation}\label{vpm}
  V_{\pm}=\frac{1}{\gamma^2}
  \e^{\pm\i\nu}\left(\delz r\pm\i\thr{}\delz t\right),
\end{equation}
if $\nu$ satisfies
\begin{equation}\label{nu}
  \delz\nu=(1+\thr{2})\delz t,\quad\delzb\nu=\chr{-2}\delzb t.
\end{equation}
Since the integrability conditions of these equations just yield the
second equation of (\ref{eom}), it will be easy to integrate them,
once we have got the general solution of the equations of motion.

Surprisingly, as in the ungauged theory, the energy-momentum tensor
has a Sugawara form
\begin{equation}
  T=\gamma^2V_+V_-,
\end{equation}
although the conformal spin-one quantities $V_\pm$ are no usual Kac-Moody
currents.

After gauging there remains the local symmetry of the ungauged WZNW
action
\begin{equation}\label{lokale-symmetrie}
t\to t+\delta\varepsilon
\end{equation}
with non-chiral Noether currents
\begin{equation}\label{strom}
J_z=\frac{1}{\gamma^2}\thr{2}\delz t,\quad
J_\zb=\frac{1}{\gamma^2}\thr{2}\delzb t.
\end{equation}
But the continuity equation
\begin{equation}\label{kontglj}
\delzb J_z+\delz J_\zb=0
\end{equation}
is equivalent, again, to the second of the equations of motion of
(\ref{eom}).

For the calculation of the symplectic structure a differential
equation of the Gelfand-Dikii type 
\begin{equation}\label{chardgl}
  y''-(\delz V_-/V_-)y'-\gamma^2Ty=0
\end{equation}
becomes important. This follows from the conserved quantities
(\ref{emt}, \ref{vpm}), and we shall show that their solutions will
usefully parametrize the general solution of the equtions of motion
(\ref{eom}).

\section{The General Solutions of the Equations\\
 of Motion and Gelfand-Dikii Equations}

The general solution of the equations of motion (\ref{eom}) is given
by \cite{MuWe}
\begin{eqnarray}\label{solution}
  \shr{2}&=&X\bar{X},\nonumber\\
  t&=&\i(B-\Bb)+\frac{\i}{2}\ln\frac{X}{\bar{X}},
\end{eqnarray}
where
\begin{eqnarray}\label{tosolution}
  X&=&A+\frac{\Bb'}{\Ab'}(1+A\Ab),\nonumber\\
  \bar{X}&=&\Ab+\frac{B'}{A'}(1+A\Ab),
\end{eqnarray}
and it is parametrized as in other non-linear theories by
arbitrary chiral (anti-chiral) functions $A(z)$, $B(z)$ ($\Ab(\zb)$,
$\Bb(\zb)$) and their derivatives $A'(z)$, $B'(z)$ ($\Ab'(\zb)$,
$\Bb'(\zb)$), respectively. This solution is invariant under internal
$\GL(2,\cc)$ transformations
\begin{eqnarray}\label{moebius}
  A&\to&T[A]=\frac{aA-b}{cA+d} ,\nonumber\\
  B&\to&T[B]=B+\ln(c
A+d),\nonumber\\
  \Ab&\to&T[\Ab]=\frac{d\Ab-c}{b\Ab+a} ,\nonumber\\
  \Bb&\to&T[\Bb]=\Bb+\ln(b\Ab+a), \\ &&\left(
\begin{array}[]{cr}
  a&-b\\ c&d
\end{array}\right)
\in \GL(2,\cc).\nonumber
\end{eqnarray}
But it does not factorize into a chiral and anti-chiral part as for
ungauged WZNW theories.  This solution allows us to integrate the 
equation for $\nu$ (\ref{nu})
\begin{equation}\label{nue-explizit}
\nu=t+\i(B+\bar B)+\i\ln(1+A\bar A)-\frac{\i}{2}\ln(1+X\bar X)+
\nu_0,
\end{equation}
and the $\GL(2,\cc)$ invariance is respected if the integration
constant $\nu_0$ transforms as
\begin{equation}\label{GL2C-nue}
\nu_0\to\nu_0-\i\ln(ad+bc).
\end{equation}

It is worth mentioning here that this general solution is
asymptotically related \cite{Buckow} to the solution of a non-abelian
Toda theory \cite{GS, Bilal} which arises by nilpotent gauging a
corresponding WZNW model.

Now it is easy to solve the Gelfand-Dikii equation (\ref{chardgl}).
We rewrite their coefficients using the general solution
(\ref{solution}, \ref{tosolution}) as
\begin{equation}\label{eit-AB}
T=\frac{1}{\gamma^2}\left(B''-B'^2-\frac{A''}{A'}B'\right),
\end{equation}
and
\begin{eqnarray}\label{V-AB}
V_+=\frac{1}{\gamma^2}\left(\frac{B''}{A'}-\frac{B'^2}{A'}-
\frac{A''B'}{A'^2}\right)\e^{\i\nu_0-2B},&&\quad
V_-=\frac{1}{\gamma^2}A'\e^{-\i\nu_0+2B},
\end{eqnarray}
and find for the Gelfand-Dikii eqution the two independent solutions
\begin{equation}\label{y12def}
y_1=\e^B,\quad\quad y_2=A\e^B.
\end{equation}
For the corresponding anti-chiral Gelfand-Dikii equation we get
\begin{equation}\label{yb12def}
\bar y_1=\e^{\bar B},\quad\quad \bar y_2=\bar A\e^{\bar B}.
\end{equation}

Thus, the general solution of the equations of motion (\ref{solution},
\ref{tosolution}) can be re\-pa\-ra\-met\-rized by the solutions of the
Gelfand-Dikii equations $y_k(z)$ and $\bar y_k(\zb)$ ($k=1,2$). A
very symmetrical expression arises for the transformed fields $u$,
$\bar u$ (\ref{lichtkegel})
\begin{equation}\label{loes-y12}
u=\frac{\bar y_1 y_1'+\bar y_2 y_2'}{y_1 y_2'-y_1'y_2},\quad
\bar u=\frac{y_1\bar y_1'+y_2\bar y_2'}{\bar y_1\bar y_2'-\bar
  y_1'\bar y_2},
\end{equation}
where for simplicity we shall restrict ourselves in the following to
the regular solutions by assuming that for finite $z$, $\bar z$
\begin{equation}\label{regulaer1}
y_1 y_2'-y_1'y_2\neq0,\quad\bar y_1 \bar y_2'-\bar y_1'\bar y_2\neq0.
\end{equation}
The $\GL(2,\cc)$ invariance (\ref{moebius}) now becomes
\begin{eqnarray}\label{GL2C-y12}
\left(\begin{array}{c}y_2\\y_1\end{array}\right)
\to
\left(\begin{array}{cc}a&-b\\c&d\end{array}\right)
\left(\begin{array}{c}y_2\\y_1\end{array}\right),\quad
\left(\begin{array}{c}\bar y_2\\\bar y_1\end{array}\right)
\to
\left(\begin{array}{cc}d&-c\\b&a\end{array}\right)
\left(\begin{array}{c}\bar y_2\\\bar y_1\end{array}\right),\\
\left(\begin{array}{cc}a&-b\\c&d\end{array}\right)\in
\GL(2,\cc).\nonumber
\end{eqnarray}

These results are useful to find the complete symplectic structure of
the theory, which is defined by the canonical Poisson brackets of the
physical fields $r(\tau, \sigma)$, $t(\tau, \sigma)$ at equal $\tau$
(writing up here and in the following the non-vanishing brackets only!)
\begin{eqnarray}\label{pk-rt}
&&\hspace{-1em}
\{r(\tau,\sigma),\dot{r}(\tau,\sigma')\}=
\gamma^2\delta(\sigma-\sigma'),\quad
\{t(\tau,\sigma),\dot{t}(\tau,\sigma')\}=
\gamma^2\coth^2\!r\;\delta(\sigma-\sigma'),\nonumber\\
&&\hspace{-1em}
\{\dot{r}(\tau,\sigma),\dot{t}(\tau,\sigma')\}=
2\gamma^2\frac{\dot{t}(\sigma,\tau)}
{\sinh r(\tau,\sigma)\cosh r(\tau,\sigma)}\delta(\sigma-\sigma').
\end{eqnarray}
According to this the canonical conjugated momenta $\pi_r$ and $\pi_t$
are 
\begin{equation}\label{impulse}
\pi_r(\tau,\sigma)=\frac{1}{\gamma^2}\dot{r},\quad
\pi_t(\tau,\sigma)=\frac{1}{\gamma^2}\thr{2}\dot{t}.
\end{equation}
The corresponding Poisson brackets of the $u$, $\bar u$ fields become
\begin{eqnarray}\label{pk-uuq}
&&\{u(\tau,\sigma),\dot{\bar u}(\tau,\sigma')\}=
\{\bar u(\tau,\sigma),\dot{u}(\tau,\sigma')\}=
2\gamma^2(1+u\bar u)\delta(\sigma-\sigma'),\nonumber\\
&&\{\dot u(\tau,\sigma),\dot{\bar u}(\tau,\sigma')\}=
2\gamma^2(\dot u\bar u-
u\dot{\bar u})\delta(\sigma-\sigma').
\end{eqnarray}
It is a non-trivial problem to calculate the different (anti-)chiral
functions of the theory in terms of the physical fields. This could be
done by solving the Gelfand-Dikii equtions. However, because of the
$\GL(2,\cc)$ invariance these functions are determined up to four
complex constants only, and we have to fix the $\GL(2,\cc)$ invariance
in order to be able to solve this initial state problem uniquely.

\section{The Solution of an Initial State Problem}

We assume that the physical fields at the 'time' $\tau_0$ take the
initial values
\begin{equation}\label{Anfangsbed}
r(\tau_0,\sigma)=r_0(\sigma),\quad t(\tau_0,\sigma)=t_0(\sigma),\quad
\dot r(\tau_0,\sigma)=r_1(\sigma),\quad
\dot t(\tau_0,\sigma)=t_1(\sigma).
\end{equation}
This also fixes the initial state of the $u$, $\bar u$ fields. Then
the $A(z)$, $B(z)$, $\Ab(\zb)$, $\Bb(\zb)$ are completely determined
through the solutions of the Gelfand-Dikii equations $y_k(z)$, $\bar
y_k(\bar z)$ (\ref{y12def}, \ref{yb12def}), in case, the $\GL(2,\cc)$
invariance (\ref{moebius}, \ref{GL2C-y12}) is fixed by four additional
initial values. But this solves the initial state problem of the two
Gelfand-Dikii equations only if we take into consideration that these
two chiral respectively anti-chiral second order differential equations
are equivalent to the four non-chiral first order differential
eqations
\begin{eqnarray}\label{dgl-erster-ord}
&&y_1'=\frac{\delz\bar u}{1+u\bar u}\left(uy_1-\bar y_2\right),\quad
y_2'=\frac{\delz\bar u}{1+u\bar u}\left(uy_2+\bar y_1\right),\nonumber\\
&&\bar y_1'=\frac{\delzb u}{1+u\bar u}
\left(\bar u\bar y_1-y_2\right),\quad
\bar y_2'=\frac{\delzb u}{1+u\bar u}
\left(\bar u\bar y_2+y_1\right),
\end{eqnarray}
which follow from the general solution (\ref{loes-y12}). 

We fix here the $\GL(2,\cc)$ invariance by the asymptotic boundary
conditions
\begin{eqnarray}\label{nebenbed}
&&\left.y_k\right|_{\sigma=-\infty}=C_k,\quad
\left.\bar y_k\right|_{\sigma=-\infty}=\bar C_k,\\
&&C_k, \bar C_k\in\cc,\quad k\in\{1,2\}.\nonumber
\end{eqnarray}
These boundary conditions solve the first order
differential equations (\ref{dgl-erster-ord}) uniquely.
Taking into consideration that the integration of (\ref{nu}) in
terms of $r$, $t$ 
\begin{eqnarray}\label{nue-integral}
\nu(\tau,\sigma)&=&t(\tau,\sigma)+
\int_{-\infty}^\sigma\d\sigma'\thr{2}\!\!(\tau,\sigma')\;
\dot t(\tau,\sigma')
\end{eqnarray}
is defined if
\begin{equation}\label{asymp-rt}
\lim_{\sigma\to\pm\infty}r(\tau,\sigma)=0,\quad
\lim_{\sigma\to\pm\infty}t(\tau,\sigma)={\rm const},
\end{equation}
respectively
\begin{equation}\label{asymp-uuq}
\lim_{\sigma\to\pm\infty}u(\tau,\sigma)=
\lim_{\sigma\to\pm\infty}\bar u(\tau,\sigma)=0,
\end{equation}
then the initial values for the derivatives are determined by eq.\ 
(\ref{dgl-erster-ord}) 
\begin{eqnarray}\label{nebenbed-abl}
&&\left.y_1'\right|_{\sigma\to-\infty}\sim-\delz\bar u\bar C_2,\quad
\left.y_2'\right|_{\sigma\to-\infty}\sim\delz\bar u\bar C_1,\nonumber\\
&&\left.\bar y_1'\right|_{\sigma\to-\infty}\sim-\delzb u C_2,\quad
\left.\bar y_2'\right|_{\sigma\to-\infty}\sim\delzb u C_1,
\end{eqnarray}
so that the solutions of the Gelfand-Dikii equations $y_k$, 
$\bar y_k$, and likewise the $A(z)$, $B(z)$, $\bar A(\zb)$, $\bar B(\zb)$,
are given uniquely as functions of the physical fields.

These boundary conditions also fix the integration constants $\nu_0$,
$\bar\nu_0$ of (\ref{nue-explizit})
\begin{equation}\label{nue0}
\nu_0=-\bar\nu_0=-\i\ln(C_1\Cb_1+C_2\Cb_2),
\end{equation}
and the conserved $V_\pm$, $\bar V_\pm$ are defined as well. 

However, in order to calculate Poisson brackets it is not necessary to
get $y_k(z)$, $\bar y_k(\zb)$ explicitely as functions of $r$, $t$,
respectively $u$, $\bar u$. We only need functional relations among
their variations.

\section{Calculation of the Symplectic Structure\\
 of the \slu{} Field Theory}

As a consequence of the solved initial state problem, the Poisson
brackets of the (anti-) chiral fields are determined uniquely by those
of the physical fields (\ref{pk-rt}, \ref{pk-uuq}). To get the Poisson
brackets of the $y_k$, $\bar y_k$ we shall calculate the variations
$\delta y_k(z)$, $\delta\bar y_k(\zb)$, e.g., as functions of the
variations $\delta u(\tau,\sigma)$, $\delta\bar u(\tau,\sigma)$,
$\delta \pi_u(\tau,\sigma)$ and $\delta\pi_{\bar u}(\tau,\sigma)$ by
varying the Gelfand-Dikii equations (neglecting the anti-chiral part
again)
\begin{equation}\label{var-y-gl}
  \delta y_k''-(\delz V_-/V_-)\delta y_k'-\gamma^2T\delta y_k=
  \delta (\delz V_-/V_-)y_k'+\gamma^2\delta Ty_k,
\end{equation}
and the initial state conditions (using again the argument of
(\ref{asymp-uuq}) that also $\delta u$, $\delta{\bar u}$ are non-zero
only in a finite but arbitrary large region)
\begin{eqnarray}\label{var-nebenbed}
&&\left.\delta y_k\right|_{\sigma=-\infty}=\delta C_k=0,\nonumber\\
&&\left.\delta y_1'\right|_{\sigma\to-\infty}=
\left.\delta y_2'\right|_{\sigma\to-\infty}=0.
\end{eqnarray}

We have to solve, finally, the initial state problem of these
inhomogeneous differential equations. For its homogeneous part we
take the solutions of the Gelfand-Dikii equations $y_k(z)$ and solve
the inhomogeneous equation by standard methods with the initial
values (\ref{var-nebenbed}). The result becomes
\begin{eqnarray}\label{var-y-sol}
\delta y_k(z)&=&\int_{-\infty}^z{\rm d}z'
\Omega(z,z')
\left(\delta(\partial V_-/V_-)(z')y_k'(z')+
\gamma^2\delta T(z')y_k(z')\right),\nonumber\\
&&\Omega(z,z')\equiv
\frac{y_1(z')y_2(z)-y_2(z')y_1(z)}
{y_1(z')y_2'(z')-y_2(z')y_1'(z')}.
\end{eqnarray}
The Poisson brackets 
\begin{eqnarray}\label{pk-y-int}
\{y_k(z),y_l(z')\}&=&
\int_{-\infty}^z{\rm d}\zs\int_{-\infty}^{z'}{\rm d}\zs'
\Omega(z,\zs)\Omega(z',\zs')\times\nonumber\\
&&
\times\big(
\{(\partial V_-/V_-)(\zs),(\partial V_-/V_-)(\zs')\}y_k'(\zs)y_l'(\zs')+
\nonumber\\
&&\quad+\gamma^2\{T(\zs),(\partial V_-/V_-)(\zs')\}y_k(\zs)y_l'(\zs')+
\nonumber\\
&&\quad+\gamma^2\{(\partial V_-/V_-)(\zs),T(\zs')\}y_k'(\zs)y_l(\zs')+
\nonumber\\
&&\quad+\gamma^4\{T(\zs),T(\zs')\}y_k(\zs)y_l(\zs')
\big)
\end{eqnarray}
can now be calculated directly from the canonical brackets of the
physical fields (\ref{pk-rt}, \ref{pk-uuq}). The result looks rather
simple after integration over $\delta$-functions
\begin{equation}\label{pk-y-exp}
\{y_k(z),y_l(z')\}=\frac{\gamma^2}{2}
\left(y_k(z)y_l(z')-y_l(z)y_k(z')\right)
\epsilon(z-z').
\end{equation}
$\epsilon(z)$ is the sign function
\begin{equation}\label{eps-def}
\epsilon(z)\equiv\left\{\begin{array}{r@{\mbox{ for }}l}
-1&z<0,\\
0&z=0,\\
1&z>0.
\end{array}\right.
\end{equation}
From this and (\ref{y12def}) the non-vanishing Poisson brackets of
$A(z)$ and $B(z)$ easily follow as
\begin{eqnarray}\label{pk-AB}
\{A(z),A(z')\}&=&\frac{\gamma^2}{2}
\left(A(z)-A(z')\right)^2\epsilon(z-z'),\nonumber\\
\{A(z),B(z')\}&=&\frac{\gamma^2}{2}
\left(A(z)-A(z')\right)\epsilon(z-z').
\end{eqnarray}

Our calculations are not finished before we have found a free-field
realization of this symplectic structure.

\section{The Canonical Free-Field Transformation}

There are several methods to find relations between $y_k(z)$, $\bar
y_k(\zb)$ and chiral, respectively anti-chiral components $\phi_k(z)$,
$\phib_k(\zb)$ of canonical free fields ($k=1,2$)
\begin{equation}\label{psi12-komp}
\psi_k(\sigma,\tau)=\phi_k(z)+\phib_k(\zb).
\end{equation}
The most easiest and straight forward approach identifies
energy-momentum tensors
\begin{equation}\label{T-phi-y}
T(z)=\left(\delz\phi_1\right)^2+\left(\delz\phi_2\right)^2
=\frac{1}{\gamma^2}
\frac{y_1''y_2'-y_1'y_2''}{y_1y_2'-y_1'y_2},
\end{equation}
and the same holds for the anti-chiral component.

Here it is appropriate to introduce complex free fields
\begin{equation}\label{psi-komplex}
\psi=\psi_1+\i\psi_2,\quad\psib=\psi_1-\i\psi_2,
\end{equation}
which factorize the components of the energy-momentum tensor
\begin{equation}\label{T-frei-komp}
T(z)=\delz\psi\delz\psib,\quad
\bar T(z)=\delzb\psi\delzb\psib.
\end{equation}
Eq.\ (\ref{psi12-komp}) gives a corresponding chiral decomposition of
$\psi$ and $\bar{\psi}$
\begin{equation}\label{psi-komp}
\psi(\sigma,\tau)=\phi(z)+\chib(\zb),\quad
\psib(\sigma,\tau)=\chi(z)+\phib(\zb),
\end{equation}
with
\begin{eqnarray}\label{phi-def}
\phi(z)=\phi_1(z)+\i\phi_2(z),&&\quad
\phib(\zb)=\phib_1(\zb)-\i\phib_2(\zb),\nonumber\\
\chi(z)=\phi_1(z)-\i\phi_2(z),&&\quad
\chib(\zb)=\phib_1(\zb)+\i\phib_2(\zb).
\end{eqnarray}
Now we assume that the free fields are local functions of the $y_k$,
$\bar y_k$, which is true for all other quantities of the theory. The
most general solution of the ansatz (\ref{T-phi-y}) parametrizes the
complex free fields as a function of $y_k$, $\bar y_k$ by eight
arbitrary complex constants $\alpha$, $\beta$, $\bar\alpha$,
$\bar\beta$, $C$, $\Cb$, $D$ and $\Db$ \cite{Mu}
\begin{eqnarray}\label{phi-y}
\phi=\frac{1}{\gamma C}\left(
\ln\frac{\alpha y_1'+\beta y_2'}{y_1 y_2'-y_1'y_2}+D\right),
&&\quad\chi=\frac{C}{\gamma}\ln\left(\alpha y_1+\beta y_2\right),
\nonumber\\
\phib=\frac{1}{\gamma\Cb}\left(
\ln\frac{\bar\alpha \yb_1'+\bar\beta \yb_2'}{\yb_1 \yb_2'-\yb_1'\yb_2}+
\Db\right),
&&\quad\chib=\frac{\bar C}{\gamma}
\ln\left(\bar\alpha \yb_1+\bar\beta \yb_2\right).
\end{eqnarray}
This solution provides, indeed, a canonical free-field transformation,
in case, the constants do not depend on physical fields. We find from
the Poisson brackets of the $y_k$, $\bar y_k$ the relations
\begin{eqnarray}\label{pk-phi}
\pk{\phi(z)}{\chi(z')}&=&-\halb\epsilon(z-z'),\nonumber\\
\pk{\phib(\zb)}{\chib(\zb')}&=&-\halb\epsilon(\zb-\zb'),\nonumber\\
\pk{\phi(z)}{\phi(z')}&=&\pk{\chi(z)}{\chi(z')}=0,\\
\pk{\phib(\zb)}{\phib(\zb')}&=&\pk{\chib(\zb)}{\chib(\zb')}=0.\nonumber
\end{eqnarray}
But these Poisson brackets follow from (\ref{phi-def}) if the $\phi_k$,
$\bar{\phi_k}$ satisfy the canonical free-field Poisson brackets
\begin{eqnarray}\label{pk-frei}
\pk{\phi_k(z)}{\phi_l(z')}&=&-\vtel\epsilon(z-z')\delta_{kl},\nonumber\\
\pk{\phib_k(\zb)}{\phib_l(\zb')}&=&-\vtel\epsilon(\zb-\zb')\delta_{kl},\\
\pk{\phi_k(z)}{\phib_l(\zb)}&=&0.\nonumber
\end{eqnarray}
This proves that the solutions of the Gelfand-Dikii equations mediate
canonical transformations between physical and free fields.

The many free parameters of (\ref{phi-y}) can be further restricted \cite{Mu}
by taking into account invariance properties of the energy-momentum
tensor, the $\GL(2,\cc)$ invariance, the chosen boundary conditions,
and, finally, by choosing special values for remaining parameters,
and the free-field relations (\ref{phi-y}) simplify to
\begin{eqnarray}\label{phi-y-spez}
\phi=\frac{1}{\gamma}
\ln\frac{y_1'}{y_1 y_2'-y_1'y_2},
&&\quad\chi=\frac{1}{\gamma}\ln y_1,
\nonumber\\
\phib=\frac{1}{\gamma}
\ln\frac{\yb_1'}{\yb_1 \yb_2'-\yb_1'\yb_2},
&&\quad\chib=\frac{1}{\gamma}
\ln\yb_1.
\end{eqnarray}

Solving these relations with respect to $y_k(z)$ and $\bar y_k(\zb)$,
we obtain the  non-local free-field realizations
\begin{eqnarray}\label{y-phi}
&&y_1(z)=\exp\left(\gamma\chi(z)\right),\nonumber\\
&&y_2(z)=\exp\left(\gamma\chi(z)\right)
\left(\intl_{-\infty}^z\d z'
\gamma\chi'(z')\exp\left(-\gamma\phi(z')-\gamma\chi(z')\right)+
\e^{2\pi\i/3}\right),\nonumber\\
&&\yb_1(\zb)=\exp\left(\gamma\chib(\zb)\right),\\
&&\yb_2(\zb)=\exp\left(\gamma\chib(\zb)\right)
\left(\intl_{+\infty}^\zb\d \zb'
\gamma\chib'(\zb')\exp\left(-\gamma\phib(\zb')-\gamma\chib(\zb')\right)+
\e^{2\pi\i/3}\right).\nonumber
\end{eqnarray}

It is easy to show that the free-field Poisson brackets yield,
consistently, the Poisson brackets of the $y_k(z)$, $\bar y_k(\zb)$,
and we could show that these results also follow from the
Gelfand-Dikii equations, in case, their coefficients are expressed in
terms of the free fields and the initial state problem is solved anew.

This proves, finally, that the non-local free-field transformations of
the physical fields $r$, $t$, respectively $u$, $\ub$ are canonical
transformations, and we can show they are one to one. 

Before collecting these transformations in local B\"acklund
transformations, we shall add a remarkable consequence of the
symplectic structure. The conserved $V_\pm(z)$ which are related to
the \slu{} coset currents satisfy non-linear Poisson brackets
\begin{eqnarray}\label{valgebra}
  \{V_\pm(\tau,\sigma),V_\pm(\tau,\sigma')\}&\!\!\!=\!\!\!&\gamma^2
  V_\pm(\tau,\sigma)V_\pm(\tau,\sigma')\epsilon(\sigma-\sigma'),\nonumber\\ 
  \{V_\pm(\tau,\sigma),V_\mp(\tau,\sigma')\}&\!\!\!=\!\!\!&-\gamma^2
  V_\pm(\tau,\sigma)V_\mp(\tau,\sigma')\epsilon(\sigma-\sigma')+
  \frac{1}{\gamma^2}\delta'(\sigma-\sigma'),\hspace{1cm}
\end{eqnarray}
which provides the Virasoro algebra, and conformal weight one for the
$V_\pm$
\begin{equation}
  \{T(\tau,\sigma),V_\pm(\tau,\sigma')\}=-\left( \delsp
  V_\pm(\tau,\sigma')\delta(\sigma-\sigma')-
  V_\pm(\tau,\sigma')\delta'(\sigma-\sigma')\right).
\end{equation}

The algebra (\ref{valgebra}) is characteristic for parafermions
\cite{FZ,BA}, which are used in the literature for the discussion of
quantum properties of gauged WZNW theories. But our intention here is
to emphasize especially the dynamics of the theory governed by the
general solution of its equations of motion.

\section{The B\"acklund Transformation of the\\{}
  \slu{} WZNW Field Theory}

Rewriting the relations (\ref{psi-komp}) in terms of (\ref{phi-y-spez})
\begin{eqnarray}\label{psi-y-spez}
\psi(\sigma,\tau)=\frac{1}{\gamma}
\ln\frac{y_1'(z)}{y_1(z) y_2'(z)-y_1'(z)y_2(z)}+\frac{1}{\gamma}
\ln\yb_1(\zb),\nonumber\\
\psib(\sigma,\tau)=\frac{1}{\gamma}
\ln\frac{\yb_1'(\zb)}{\yb_1(\zb) \yb_2'(\zb)-\yb_1'(\zb)\yb_2(\zb)}+
\frac{1}{\gamma}\ln y_1(z)
\end{eqnarray}    
using (\ref{loes-y12}), and eliminating the functions $y_k(z)$, 
$\bar y_k(\zb)$ and their derivatives gives the B\"acklund
transformation between the physical and the free fields
\begin{equation}\label{bl}\hspace{-0.5ex}
\begin{array}{ll}
\displaystyle
\delz u=\frac{\gamma}{2}\;
\e^{\gamma\psi}\left(P+\sqrt{P^2-4Q}\right)\delz\psi,&
\displaystyle
\hspace{0ex}\delzb u=\frac{\gamma}{2}\;
\e^{\gamma\psi}\left(P-\sqrt{P^2-4Q}\right)\delzb\psi,\nonumber\\[3ex]
\displaystyle
\delz\bar{u}=\frac{\gamma}{2}\; 
\e^{\gamma\bar{\psi}}\left(P-\sqrt{P^2-4Q}\right)\delz\bar{\psi},&
\displaystyle
\hspace{0ex}\delzb\bar{u}=\frac{\gamma}{2}\;
\e^{\gamma\bar{\psi}}\left(P+\sqrt{P^2-4Q}\right)\delzb\bar{\psi}.
\end{array}\hspace{-2.5ex}
\end{equation}
$P$ and $Q$ are 
\begin{equation}\label{pq}
P(u,\bar{u},\psi,\bar{\psi})\equiv
u\,\e^{-\gamma\psi}+\bar{u}\,\e^{-\gamma\bar{\psi}}+
\e^{-\gamma\psi-\gamma\bar{\psi}},\quad
Q(u,\bar{u},\psi,\bar{\psi})\equiv
(1+u\bar{u})\;\e^{-\gamma\psi-\gamma\bar{\psi}}.
\end{equation}
The integrability conditions for the fields $u$ and $\ub$ 
\begin{eqnarray}\label{int-bed-uuq}
&&0=\delzb\delz u-\delz\delzb u=
\gamma\e^{\gamma\psi}\sqrt{P^2-4Q}\ \ \delz\delzb\psi\nonumber\\
&&0=\delzb\delz\ub-\delz\delzb\ub=
-\gamma\e^{\gamma\psib}\sqrt{P^2-4Q}\ \ \delz\delzb\psib
\end{eqnarray}
correspond to the free field equations whereas the integrability
conditions for the free fields just give the equations of motion for
the fields  $u$ and $\ub$
\begin{eqnarray}\label{int-bed-psi}
0=\delzb\delz\psi-\delz\delzb\psi=
-\frac{1}{1+u\ub}\sqrt{P^2-4Q}
\left(\delz\delzb u-\frac{\ub\delz u\delzb u}{1+u\ub}\right),\nonumber\\
0=\delzb\delz\psib-\delz\delzb\psib=
-\frac{1}{1+u\ub}\sqrt{P^2-4Q}
\left(\delz\delzb\ub-\frac{u\delz\ub\delzb\ub}{1+u\ub}\right).
\end{eqnarray}

This completes the general discussion of the \slu{} gauged WZNW field
theory, and we expect that the canonical free-field transformation
prepares the theory for a canonical quantization.

\section{Conclusion}

We have given a proof that any gauged WZNW model is an integrable
conformal theory, which was known so far for nilpotent gauging. We
have, furthermore, completely solved the \slu{} theory and calculated
its symplectic structure by solving an initial state problem of a
Gelfand-Dikii type equation, instead of suggesting this structure
only. On this basis we have derived a non-local canonical free-field
transformation of the non-linear fields of the theory and we have,
finally, collected these transformations in a local B\"acklund
transformation.

The calculations were done for a field-theoretic situation with
asymptotic boundary conditions. Similar calculations for the periodic
case, which include zero modes, will be published elsewhere.
 
Although the classical solution of the \slu{} gauged WZNW model bears
strong resemblance to Liouville or Toda theories its quantum structure
might be different because its energy-momentum tensor does not have a
`central charge' term classically. But an arising dilaton might
compensate for these differences. Thus, an exact quantization of the
\slu{} theory on this basis will be a challange, in particular, with
respect to the space-time black hole interpretation of the theory.

\section*{Acknowledgement}
We would very much like to thank C.-J. Biebl and Ch. Ford for
reading the manuscript and for discussions.


\begin{thebibliography}{77}
  \bibitem{Witten} E. Witten, {\it Phys. Rev.} {\bf D44} (1991) 314.
 \bibitem{DVV} R. Dijkgraaf, E. Verlinde, H. Verlinde, {\it Nucl.
    Phys.} {\bf B371} (1992) 269.
  \bibitem{Tseytlin} A. A. Tseytlin, {\it Nucl. Phys.} {\bf
    B399} (1993) 601.
  \bibitem{Buscher} T.H. Buscher, {\it Phys.Lett.} {\bf B201} (1988)
  466.
  \bibitem{Mu} U. M{\"u}ller, thesis 1998, written in German, unpublished.
  \bibitem{BA} K. Bardakci, M. Crescimanno, E. Rabinovici, {\it Nucl.
    Phys.} {\bf 344} (1990) 344.
  \bibitem{MuWe} U. M{\"u}ller and G.Weigt, {\it Phys. Lett.} 
  {\bf B400} (1997) 21.
  \bibitem{Balog} J. Balog, L. Feher, P. O'Raifearthaigh, A. Wipf,
     {\it Phys.Lett.} {\bf B227} (1989); {\it Annals Phys.} {\bf 203}
     (1992) 269.
  \bibitem{GS} J.-L. Gervais, M. Saveliev, {\it Phys. Lett.} {\bf
    B286} (1992) 271.
  \bibitem{GKO} P. Goddard, A. Kent, D. Olive, {\it Phys. Lett.}
  {\bf B152} (1985) 88.
  \bibitem{Callan} C.G. Callan, D. Friedan, E.J. Martinec, and 
   M.J. Perry, {\it Nucl. Phys.} {\bf B262} (1985) 593.
  \bibitem{WZNW} J. Wess, B. Zumino, {\it Phys. Lett.} {\bf B37}
    (1986) 78; S. P. Novikov, {\it Sov. Math. Dokl.} {\bf 37} (1982) 3;
    E. Witten, {\it Comm. Math. Phys.} {\bf 92} (1984) 455. 
  \bibitem{Buckow} U. M{\"u}ller and G.Weigt, in Proceedings of the
  30th International Symposium   Ahrenshoop, 1996, Buckow, Germany, p.
  328.
  \bibitem{Bilal} A. Bilal, {\it Nucl. Phys.} {\bf B422} (1994) 258. 

  \bibitem{FZ} V.A. Fateev and A.B. Zamolodchikov,
  {\it Zh.Eksp.Teor.Fiz.} {\bf 89} (1985)380.
\end{thebibliography}
\end{document}